\documentclass[physics,article,submit,pdftex,moreauthors]{Definitions/mdpi} 

\preto{\abstractkeywords}{\nolinenumbers}

\newcommand{\be}{\begin{equation}}
\newcommand{\ee}{\end{equation}}
 
 \definecolor{BrickRed}{cmyk}{0,0.89,0.94,0.28}
\definecolor{MidnightBlue}{cmyk}{0.98,0.13,0,0.43}
\definecolor{DarkGreen}{rgb}{0,0.7,0.1}

\usepackage{graphicx}
\usepackage{verbatim}
\usepackage{amsfonts} 
\usepackage{amsmath}
\usepackage{mathbbol}
\usepackage{amssymb}   
\usepackage{color}
\usepackage{bm}

\firstpage{1} 
\pubvolume{1}
\issuenum{1}
\articlenumber{0}
\pubyear{2023}
\copyrightyear{2023}
\datereceived{ } 
\daterevised{ } 
\dateaccepted{ } 
\datepublished{ } 
\hreflink{https://doi.org/} 

\Title{Surface Scattering Expansion of the Casimir-Polder Interaction for Magneto-dielectric Bodies: Convergence Properties for Insulators, Conductors and Semiconductors}

\TitleCitation{Title}

\Author{Giuseppe Bimonte $^{1,\dagger}$ and  Thorsten Emig $^{2}$}

\address{%
$^{1}$ \quad  Dipartimento di Fisica E. Pancini, Universit\`{a} di
Napoli Federico II, Complesso Universitario
di Monte S. Angelo,  Via Cintia, I-80126 Napoli, Italy; giuseppe.bimonte@na.infn.it\\
$^{\dagger}$ \quad  INFN Sezione di Napoli, I-80126 Napoli, Italy\\
$^{2}$ \quad  Laboratoire de Physique
Th\'eorique et Mod\`eles Statistiques, CNRS UMR 8626,
Universit\'e Paris-Saclay, 91405 Orsay cedex, France}

\corres{Correspondence:  giuseppe.bimonte@na.infn.it}

\abstract{Fluctuation induced forces are a hallmark of the interplay of fluctuations
and geometry. We recently proved the existence of a multi-parametric family of exact representations of
Casimir and Casimir-Polder  interactions between bodies of arbitrary shape and material
composition, admitting a multiple scattering expansion (MSE) as a sequence of inter- and intra-body
multiple wave scatterings [G.~Bimonte, T.~Emig, Phys. Rev. A 108, 052807 (2023)]. The approach requires no
knowledge of the scattering amplitude (T-matrix) of the bodies. Here we investigate the convergence properties of the
MSE for the Casimir-Polder interaction of a polarizable particle with a macroscopic body. We consider representative materials from different classes, such as insulators, conductors and semiconductors. 
Using a sphere and a cylinder as benchmarks, we demonstrate that the MSE can be used to 
efficiently and accurately compute the Casimir-Polder interaction for bodies with smooth surfaces.}


\begin{document}

 
\section{Introduction}

Following the seminal  work of Casimir, who discovered that two discharged perfectly conducting parallel plates  at zero temperature attract each other with a force originating from quantum fluctuations of the electromagnetic (em) field  \cite{casimir}, Lifshitz successfully used  the then new field of fluctuational electrodynamics to compute the Casimir force between two parallel, infinite surfaces of dispersive and dissipative dielectric bodies at finite temperature \cite{lifshitz}. By taking the dilute limit for one of the two bodies, Lifshitz could also compute the Casimir-Polder (CP) force between a small polarizable particle and a planar surface.  Lifshitz's results remained unsurpassed for a long time, because it was not clear how to extend his computation  beyond simple planar surfaces.    Computing the Casimir and Casimir-Polder (CP) interactions in non-planar geometries is  in fact a notoriously difficult problem, due to the collective and non-additive character of dispersion forces.  For a
long time the only method to estimate dispersion forces in non-planar setups was the Derjaguin additive approximation \cite{derjaguin},  the so-called Proximity Force Approximation (PFA), which expresses the Casimir force between two non-planar surfaces as the sum of the forces between  pairs of small opposing planar portions of the surfaces. Because of  its simplicity, the PFA is still widely used to interpret modern experiments with curved bodies. For reviews, see \cite{parsegian,galina,buhmann1,rodriguez,woods,bimonte2017,bimonte2022}. 

A significant step forward in the study of curved surfaces was made in the seventies of last century by Langbein, who used scattering methods to study the Casimir interaction between spheres and cylinders \cite{langbein}. The remarkable work of Langbein went largely unnoticed, and was quickly forgotten. A new wave of strong interest in the problem arose at the beginning of this century, spurred by modern precision experiments on the Casimir effect \cite{lamoreaux,mohideen,chan,bressi,Decca:2003yb,Munday:2009xw,Sushkov:2011ik,Tang:2017kz,Bimonte:2016cr}.   
The intense theoretical efforts that were put forward culminated in the discovery of the scattering formula \cite{emig2007,kenneth2008,rahi2009}, initially devised for non-planar mirrors \cite{Genet03,lambrecht}, which expresses the interaction between dielectric bodies in terms of their scattering amplitude, known as T-operator. While this approach has enabled most of recent theoretical progress, the T-operator is known only for highly symmetric bodies, such as sphere and cylinder, or for a few perfectly conducting shapes \cite{maghrebi}. 
Remarkably enough, it has been found  that the scattering formula can be computed exactly for the  sphere-plate and the sphere-sphere systems, for Drude conductors in the high temperature limit \cite{bimonte2012ter,Ingold2021}.  By improved numerical methods, the scattering formula for a dielectric sphere and a plate at finite temperature can be computed  with high precision also for experimentally relevant small separations \cite{hartmann2017}. We note, however, that the precision of current experiments using the simple sphere-plate geometry has not yet reached the point where deviations from the PFA can be observed. 

As we said above, the practical use of the scattering approach is limited to the few simple shapes for which the scattering  amplitude is known. A more fundamental limitation of the scattering approach is that interlocked geometries evade this method due to lack of convergence of the mode expansion \cite{Wang2021}.  The necessity of  theoretical formulations for a precise force computation in  complex geometries  has become urgent lately, because recent experiments using micro-fabricated surfaces \cite{Banishev:2013zp,Intravaia:2013yf,Wang2021} have shown indeed large deviations from the PFA.   Theoretical progress has been made for the special case of  dielectric rectangular gratings, by  using  a generalization of the Rayleigh expansion in \cite{marachevsky2008,marachevsky2014,marachevsky2020}. On a different route,  a general semi-analytical approach  has been devised  for  gently curved surfaces,  for which  a gradient expansion can be used to obtain first order curvature corrections to the proximity force approximation both for the Casimir force \cite{fosco,bimonte2012,bimonte2012bis,bimonte2017bis} and for the CP interaction
\cite{bimonte2014,bimonte2015}.    

A  breakthrough occurred in 2013 \cite{reid2013}  when it was shown  that surface integral-equations methods \cite{chew,volakis}, that have been used for a long time  in  computational electromagnetism,  can be also used  to compute, at least in principle, Casimir interactions  for  arbitrary arrangements of any number of (homogeneous) magneto-dielectric bodies of any  shape.  The  formulation in  \cite{reid2013}  expresses  Casimir   forces and energies  as traces of certain expressions involving a  surface operator, evaluated along the imaginary frequency axis. 
The surface operator consists of  linear combinations with constant coefficients of free Green tensors of the em field of $N+1$ {\it homogeneous} infinite media,  having the  permittivities  of the $N$ bodies, and of the medium surrounding them.   A potential problem with the approach of \cite{reid2013}  is that the expression for the Casimir interaction contains the inverse of the surface operator, that has to be computed numerically by replacing the continuous surfaces with a suitable discrete mesh.  This operation replaces the surface operator by a large matrix  whose elements  involve double surface integrals of the free Green tensors over all pairs of small surface elements composing the mesh.   The generation of the matrix  is time consuming, because of the strong inverse-distance cubed singularity of the surface operator in the coincidence limit.  In addition to that, the size of the non-sparse matrix for sufficiently fine meshes can quickly
exceed the memory-usage limit, preventing its numerical inversion.

 Inspired by older work by Balian and Duplantier on the Casimir effect for perfect conductors \cite{balian1977,balian1978}, we have recently derived a multiple scattering expansion (MSE)  of Casimir and CP interactions for magneto-dielectric bodies of arbitrary shape \cite{emig2023,bimonte2023}.  Similar to Ref.~\cite{reid2013}, in our approach the interactions have the form of traces of expressions involving the inverse of a surface operator $\mathbb{M}({\rm i}\xi)$, evaluated along the imaginary frequency axis. A crucial difference with respect to Ref.~\cite{reid2013} is that our kernel $\mathbb{M}$ has the form of a Fredholm surface integral operator  of the second kind,
 \be
 \mathbb{M}= \mathbb{I}-\mathbb{K}\;. \label{fred}
 \ee
 The Fredholm form  implies that  the inverse $\mathbb{M}$  can be computed as a power (Neumann) series
 \be
 \mathbb{M}^{-1}= \mathbb{I} + \mathbb{K}+  \mathbb{K}^2 + \dots ,
 \ee 
 which converges provided that the spectral radius of $\mathbb{K}$ is less than one.
 Hence, we  obtain  an expansion of   Casimir and CP interaction in powers of $\mathbb{K}$, which can be physically interpreted as an expansion  in the number of  scatterings off the surfaces of the bodies. Specifically, the MSE  has the form of an iterated  series of surface integrals of elementary functions, running over the surfaces of the bodies.  We note that a particular choice of free coefficients in the kernel $\mathbb{K}$ exists, such that it has a weak $1/|{\bf u}-{\bf u}'|$ singularity. This should  simplify and accelerate numerical evaluations on a mesh.   An additional advantage  implied by the MSE, if implemented on a mesh, is that one does not need to store the  matrix for $\mathbb{K}$ in  memory, since its elements can be  computed at the moment of performing the matrix multiplication.  

 In \cite{emig2023} we showed that only a few terms of the MSE are sufficient to obtain a fairly accurate estimate of the Casimir energy between a Si wedge and a Au plate.  The purpose of the present work  is to investigate the convergence properties  of the MSE for the CP interaction between a polarizable isotropic particle and a dielectric body. We use as benchmarks two shapes that can be solved exactly by using the scattering approach, namely a sphere or a cylinder.  We  consider  different types of materials for the sphere and the cylinder, in order to see how the material properties of the bodies affect the rate of convergence of the MSE. We demonstrate that in all cases the MSE converges fast and uniformly with respect to the particle-surface separation. Since there is no reason to expect that the convergence properties of the MSE will be any different for bodies that are smooth deformations of a sphere or a cylinder, we argue that the findings of this work imply that the MSE can be used to efficiently compute the CP interactions for compact and non-compact dielectric bodies with smooth surfaces of any shape.


\section{MSE of the scattering Green tensor}

Consider a collection of $N$ magneto-dielectric bodies with surfaces $S_{\sigma}$ ($\sigma=1,\cdots N$), characterized by frequency dependent electric and magnetic permeabilities $\epsilon_{\sigma}(\omega)$ and $\mu_{\sigma}(\omega)$, respectively, embedded in a homogeneous medium with permittivities  $\epsilon_{0}(\omega)$ and $\mu_{0}(\omega)$, and let $\mathbb{G}$ is the $N$-body EM Green tensor. We define the scattering Green tensor ${\Gamma}({\bf r},{\bf r}')$ as
\be
{\Gamma}({\bf r},{\bf r}')=\mathbb{G}({\bf r},{\bf r}')-\mathbb{G}_0({\bf r},{\bf r}')
\ee 
where  $\mathbb{G}$ is the $N$-body EM Green tensor and $\mathbb{G}_0$ is the empty space Green tensor for a homogenous medium with contrast $\epsilon_0$, $\mu_0$ (see App.~E of \cite{bimonte2023} for the definition of $\mathbb{G}_0$).  Physically, ${\Gamma}({\bf r},{\bf r}')$ describes the {\it modification} of the EM field at position ${\bf r}$ due to the presence of the bodies, i.e. the scattered field  generated by a source $({\bf J}({\bf r}'),{\bf M}({\bf r}'))$ at position ${\bf r}'$ outside the bodies. In the surface-integral approach, one imagines that the scattered field at ${\bf r}$ is radiated  by fictitious tangential electric and magnetic  surface currents  $({\bf j}_{\sigma}({\bf u}), {\bf m}_{\sigma}({\bf u}))$ located at points ${\bf u}$ on the surfaces of the bodies.
One can show  \cite{bimonte2023} that the surface currents satisfy the following set of Fredholm integral equations of the second kind,
\begin{equation}
\label{eq:Fredholm_currents}
\sum_{\sigma'=1}^N\int_{S_{\sigma'}}\!\!ds_{{\bf u}'} \,\left[\mathbb{{I}} - \mathbb{K}_{\sigma\sigma'}({\bf u},{\bf u}')\right] \big(\begin{smallmatrix} {\bf j}_{\sigma'} \\ {\bf m}_{\sigma'} \end{smallmatrix} \big) ({\bf u}')
= \!\int d{\bf r} \,\mathbb{M}_{\sigma}({\bf u},{\bf r})\big(\begin{smallmatrix} {\bf J} \\ {\bf M}\end{smallmatrix}\big) ({\bf r})\;.
\end{equation}
In the above Equation, $\mathbb{K}_{\sigma\sigma'}({\bf u},{\bf u}')$ and $\mathbb{M}_{\sigma\sigma'}({\bf u},{\bf u}')$  denote the   surface scattering operator (SSO)
\begin{equation}
\label{eq:1}
\mathbb{K}_{\sigma\sigma'}({\bf u},{\bf u}') = 2 \mathbb{P} (\mathbb{C}^{i}_\sigma+\mathbb{C}^{e}_\sigma)^{-1} {\bf n}_\sigma({\bf u}) \times  \left[  \delta_{\sigma\sigma'} \mathbb{C}^{i}_\sigma  \mathbb{G}_\sigma({\bf u},{\bf u}')- \mathbb{C}^{e}_\sigma \mathbb{G}_0({\bf u},{\bf u}')\right] \, , \quad \mathbb{P} = \big(\begin{smallmatrix} 0 & -1 \\ 1 & 0 \end{smallmatrix} \big)
\end{equation}
and the operator
\begin{equation}
\label{eq:3}
\mathbb{M}_\sigma({\bf u},{\bf r}) =  -2 \mathbb{P} (\mathbb{C}^{i}_\sigma+\mathbb{C}^{e}_\sigma)^{-1} \mathbb{C}^{e}_\sigma\, {\bf n}_\sigma({\bf u}) \times  \mathbb{G}_0({\bf u},{\bf r}) \, .
\end{equation}
where ${\bf n}_\sigma({\bf u}) $ is the outward unit normal vector at point ${\bf u}$,  and $\delta_{\sigma\sigma'}$ is the Kronecker delta.  The  operators $\mathbb{K}$ and $\mathbb{M}$ act on electric and magnetic tangential surface fields at ${\bf u}'$ . The action of ${\bf n}_\sigma({\bf u}) \times$ on the $3 \times 3$ matrices $\mathbb{G}^{(pq)}_{\sigma}$ and $\mathbb{G}^{(pq)}_{0}$ ($p,q \in \{ E,H\}$) are respectively defined by $({\bf n}_\sigma({\bf u}) \times \mathbb{G}^{(pq)}_{\sigma}){\bf v} \equiv {\bf n}_\sigma({\bf u}) \times (\mathbb{G}^{(pq)}_{\sigma} {\bf v})$ and  $({\bf n}_\sigma({\bf u}) \times \mathbb{G}^{(pq)}_{0}){\bf v} \equiv {\bf n}_\sigma({\bf u}) \times (\mathbb{G}^{(pq)}_{0} {\bf v})$,  for any vector ${\bf v}$.  We note that $\mathbb{K}$ and $\mathbb{M}$ depend on  $4N$  arbitrary coefficients, which must form $2N$ invertible diagonal $2\times 2$ matrices $\mathbb{C}^{i}_\sigma$, $\mathbb{C}^{e}_\sigma$. Uniqueness of the surface currents implies that for all choices of the these  coefficients, the currents solving Eq.~(\ref{eq:Fredholm_currents}) for given sources  $({\bf J}({\bf r}'),{\bf M}({\bf r}'))$  are the same.

The previous considerations imply that the  scattering Green tensor has the  representation
\begin{equation}
\label{eq:Gamma}
{\Gamma}({\bf r},{\bf r}')=
\int_S ds_{\bf u} \int_S ds_{{\bf u}'}\,  \mathbb{G}_0({\bf r},{\bf u}) (\mathbb{I}-\mathbb{K})^{-1}({\bf u},{\bf u}') \mathbb{M}({\bf u}',{\bf r}')
\end{equation}
where the integration extends over all surfaces $S_\sigma$ and a summation over all  surface labels $\sigma$ is understood. The existence of a MSE follows from the Fredholm type of the operator $(\mathbb{I}-\mathbb{K})^{-1}$ that permits an expansion in powers of $\mathbb{K}$.

\section{Casimir-Polder energy of a polarizable particle and a magneto-dielectric body}

\begin{figure}[H]
\includegraphics[width=0.55\textwidth]{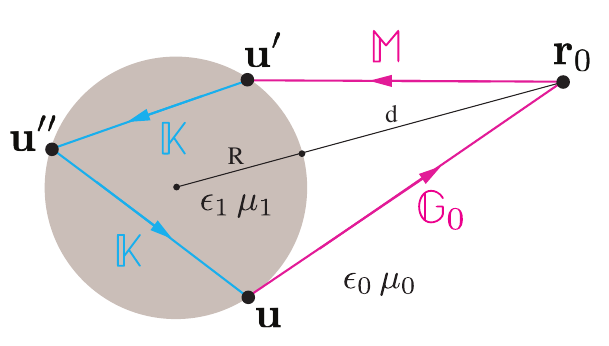}
\caption{Configuration of a dielectric body (sphere, cylinder) of radius $R$ with electric and magnetic permeabilities $\epsilon_1$, $\mu_1$, interacting with a polarizable particle at position ${\bf r}_0$ outside the object at a distance $d$ from the surface. The electric and magnetic permeabilities of the surrounding medium are $\epsilon_0$, $\mu_0$. Shown are also the operators of the multiple scattering expansion, see text for details. 
\label{Fig1}}
\end{figure}

We consider now the Casimir-Polder interaction between a polarizable particle and a magneto-dielectric body, see Fig.~\ref{Fig1}. We assume that the particle is characterized by a frequency dependent electric polarizability tensor ${\bm \alpha}(\omega)$ and a magnetic polarizability tensor ${\bm \beta}(\omega)$. The classical energy of an induced dipole is then given by
\begin{equation}
\label{eq:inducedDipole}
{E}_\text{cl} = -\frac{1}{2} \sum_{i,j=1}^{3} \left[ \alpha_{ij} E_i E_j + \beta_{ij} H_i H_j\right] \, .
\end{equation}
Using the fluctuation-dissipation theorem, this expression is averaged over EM field fluctuations. After removing a divergent contribution from empty space, the Casimir Polder energy is expressed in terms of the scattering Green tensor as
\begin{equation}
\label{eq:CPenergy}
{E}_\text{CP} = - 4\pi k_B T \, \sideset{}{'}\sum_{n=0}^\infty  \kappa_n \, \sum_{i,j=1}^{3} \left[ \alpha_{ij} ({\rm i} \, \xi_n) {\Gamma}^{(EE)}_{ij}({\bf r}_0,{\bf r}_0;\kappa_n)+ \beta_{ij} ({\rm i} \, \xi_n) {\Gamma}^{(HH)}_{ij}({\bf r}_0,{\bf r}_0;\kappa_n) \right] \, ,
\end{equation}
where $k_B$ is Boltzmann constant, $T$ is the temperature,  $\xi_n= 2 \pi n k_B T/\hbar$, $n=0,1,2\cdots$ are the  Matsubara frequencies, $\kappa_n=\xi_n/c$, the prime in the sum indicates that the $n=0$ terms has to be taken with a weight of $1/2$, and  ${\bf r}_0$ is the particle's position. 
Substitution of ${\Gamma}$ from Eq.~(\ref{eq:Gamma}) yields the interaction energy of the particle with the body in terms of the SSO.
This energy can be computed by a MSE with respect to the number of scatterings at the surface of the body. It is instructive to consider explicitly the first terms of the scattering expansion of the Casimir-Polder energy, assuming for simplicity that the electric polarizability of the particle is isotropic $\alpha_{ij}=\alpha \,\delta_{ij}$ , and that its magnetic polarizability $\beta$ is negligible, yielding
\begin{align}
{E}_\text{CP} &= - 4\pi k_B T \, \sideset{}{'}\sum_{n=0}^\infty  \kappa_n \,  \alpha({\rm i} \, \xi_n)\, \left\{ \sum_{p={E,H}} \int_S ds_{\bf u}  {\rm tr} \left[\mathbb{G}_{0}^{(E p)}({\bf r}_0,{\bf u}; \kappa_n)    \mathbb{M}^{(p E)}({\bf u},{\bf r}_0; \kappa_n) \right] \right. \nonumber \\
&\left.+\sum_{p,q={E,H}} \int_S ds_{\bf u}  \int_S ds_{{\bf u}'}  {\rm tr} \left[\mathbb{G}_{0}^{(E p)}({\bf r}_0,{\bf u}; \kappa_n)   \mathbb{K}^{(p q)}({\bf u},{\bf u}'; \kappa_n)   \mathbb{M}^{(q E)}({\bf u}',{\bf r}_0; \kappa_n) \right] \right\}+\cdots \label{CPser}
\end{align}
where ${\rm tr}$ denotes a trace over tensor spatial indices. Recalling that the kernels $\mathbb{K}$ and $\mathbb{M}$ are combinations of free-space Green tensors $\mathbb{G}_{0}$ and $\mathbb{G}_{\sigma}$, and that the latter are elementary functions, we see from the above equation that the CP energy is expressed in terms of iterated integrals of elementary functions extended on the surface $S$ of the body. Since for imaginary frequencies the Green tensors decay exponentially with distance, Eq. (\ref{CPser}) makes evident the intuitive fact that the points of the surface that  are closest to the particle dominate the interaction.  However, for the classical term $n=0$ the Matsubara frequency vanishes, and the operators decay only according to a power law.

\section{Equivalent formulations of the SSO}
\label{sec:equiform}

With different interior coefficient matrices $\mathbb{C}^{i}_\sigma$ and exterior coefficient matrices $\mathbb{C}^{e}_\sigma$ the SSO form an equivalence class of operators in the sense that Eq.~(\ref{eq:Fredholm_currents}) yields the same surface currents for a given external source for all coefficients, as long as neither the interior nor the exterior matrices vanish for any $\sigma$, and the sum $\mathbb{C}^{i}_\sigma+\mathbb{C}^{e}_\sigma$ is invertible. Consequently, the scattering Green tensor and the CP energy must be also independent of the choice made for the coefficients. The surface currents and the Casimir energy at any {\it finite} order of the MSE, however, in general do depend on the chosen coefficients, and hence does the rate of convergence of the MSE. This remarkable property provides an effective method to optimize convergence for different permittivities and even frequencies by suitable adjustment of coefficients. 
Among the infinitely many choices there are two which we consider important to discuss explicitly:

\noindent
{\bf (C1)}  In general, the SSO has a leading singularity that diverges as $1/|{\bf u}-{\bf u}'|^\gamma$ with $\gamma=3$ when the two surface positions ${\bf u}$, ${\bf u}'$ approach each other. 
There exists a choice of coefficients \cite{muller}, however, for which the singularity is reduced to a weaker divergence with exponent $\gamma=1$. 
The coefficient matrices are 
\begin{equation}
\mathbb{C}^{i}_\sigma={\rm diag}(\epsilon_\sigma,\mu_\sigma)\, , \quad
\mathbb{C}^{e}_\sigma={\rm diag}(\epsilon_0,\mu_0) \, .\label{C1}
\end{equation}
The corresponding surface operator $ \mathbb{K}$  has unique mathematical properties  (see Sec.~VI of \cite{bimonte2023}). 

\noindent
{\bf (C2)} A fully asymmetric, material independent choice of coefficient matrices is 
\begin{equation}
\mathbb{C}^{i}_\sigma={\rm diag}(1,0)\, , \quad \mathbb{C}^{e}_\sigma={\rm diag}(0,1) \,.\label{C2}
\end{equation}
 
\section{Results and Discussion}

The MSE of the CP energy Eq. (\ref{CPser}) converges if all eigenvalues of the SSO $\mathbb{K}$ are less than one in modulus, which we call the boundedness property. Unfortunately, we have not been able to derive a general bound on the eigenvalues of $\mathbb{K}$. However, we could prove \cite{bimonte2023} the boundedness property for the choice (C1) of the coefficients [see Eq.~(\ref{C1})] in the asymptotic limit of infinite frequencies for bodies of any shape. For compact bodies, the boundedness property holds also in the static limit $\kappa=0$. For the special case of  
perfect conductors of compact shape, the boundedness property  was proven long time ago at all frequencies \cite{balian1977}.

While having a  proof of convergence of the MSE is clearly desirable, from the practical point of view it is more important to know if convergence is fast enough that the first few terms of the MSE  provide a  good approximation to the complete series. Given the current status of experiments, getting the CP energy with an error less than say a percent would be good enough. To investigate this problem, we thought of using as a benchmark the CP interaction of a particle with a body for which the scattering amplitude (T-matrix) is exactly known, and then to verify in such a setup the rate of convergence of the MSE expansion to the exact energy.  We chose to study a dielectric sphere and a dielectric cylinder. We consider three different materials, representing a conductor (gold), a semiconductor (silicon) and an insulator (polystyrene). Since these materials  have widely different permittivities,   we can  check how the rate of convergence of the MSE is affected by the magnitude of the permittivity. We have compared the rate of convergence of the MSE for the two choices (C1) and (C2)  of the free coefficients that enter in the definition of the SSO.   We shall denote by ${\rm MSE}_k$, $k=0,1,\ldots$ the estimate of the CP energy corresponding to including up-to $k$ powers of $\mathbb{K}$  in the MSE of Eq. (\ref{eq:CPenergy}).

\subsection{Materials}
 In our computations we used for the permittivities of the materials the  expressions 
\begin{eqnarray}
\epsilon_{\rm Au}({\rm i} \, \xi_n)&=& 1+\frac{\Omega_p^2}{\xi (\xi+\gamma)}+ \sum_j \frac{f_j}{\omega_j^2+g_j \xi+\xi^2}\;,\\
\epsilon_{\rm Si}({\rm i} \, \xi_n)&=& \epsilon_{\infty}^{(\rm Si)} + \frac{\epsilon^{(\rm Si)} _0- \epsilon^{(\rm Si)} _{\infty}}{ 1+ \xi^2/ \omega_{\rm UV}^2}\;,\\
\epsilon_{\rm polystyrene}({\rm i} \, \xi_n)&=&1+\sum_j \frac{f_j}{\omega_j^2+g_j \xi+\xi^2}\;,
\end{eqnarray}
where $\Omega_p=9$ eV$/\hbar$,  $\gamma=0.035$ eV$/\hbar$, $\epsilon_{\infty}^{(\rm Si)} =1.035$, $\epsilon^{(\rm Si)} _0=11.87$, $ \omega_{\rm UV}= 4.34$ eV$/\hbar$, and the oscillator parameters $\omega_j,f_j, g_j$ for Au and polystyrene are listed in Tables (\ref{tabAu}) and (\ref{tabpoly}) respectively. We assume that the particle's polarizability $\alpha$ is frequency independent.

\subsection{CP energy for a sphere}

The scattering approach yields for the CP interaction energy of a polarizable particle at distance $d$ from the surface of a sphere of radius $R$ in vacuum ($\epsilon_0=\mu_0=1$)  the result
\begin{eqnarray}
\label{eq:E_sphere}
{E}^\text{(exact)}_\text{CP} &=&\frac{k_B T}{a^2} \sideset{}{'}\sum_{n=0}^\infty  \kappa_n \,  \alpha({\rm i} \, \xi_n) \sum_{l=1}^{\infty} (2 l +1) \\
& \times &  \left\{T^{\rm HH}_{l}  ({\rm i} \, \xi_n)  {\cal K}_l^2 (\kappa_n a) -
T^{\rm EE}_l ({\rm i} \, \xi_n) \left[ {\cal K}_l^{'2} (\kappa_n a) + \frac{l(l+1)}{\kappa_n^2 a^2} {\cal K}_l^2 (\kappa_n a)  \right]\right\}\nonumber \;,
\end{eqnarray}
where $a=R+d$,  $l$ is the multipole index,  ${\cal K}'_l(x)= d {\cal K}_l /dx$,  ${\cal K}_l (x) = x k_l(x)$,  $k_l(x) = \sqrt{\frac{2}{\pi x}} K_{l+1/2}(x)$  is the modified spherical Bessel function of the third kind, and $T^{{\rm HH}}_l , T^{\rm EE}_l $ are the T-matrix elements (Mie coefficients) of the sphere,
\begin{eqnarray}
T^{\rm HH}_{l} ({\rm i} \xi)&=& \frac{ \sqrt{\mu/\epsilon}\, {\cal I}_l (\sqrt{\epsilon \mu}\kappa R) \; {\cal I}'_l (\kappa R) -   {\cal I}'_l (\sqrt{\epsilon \mu}\kappa R)  \;{\cal I}_l (\kappa R)  }
{{\cal K}_l (\kappa R) \; {\cal I}'_l ( \sqrt{\epsilon \mu}\kappa R) - \sqrt{\mu/\epsilon} \; {\cal I}_l (\sqrt{\epsilon \mu}\kappa R) \; {\cal K}'_l (\kappa R) } \;,\\
T^{\rm EE}_{l} ({\rm i} \xi)&=& \frac{ \sqrt{\epsilon/\mu}\, {\cal I}_l (\sqrt{\epsilon \mu}\kappa R) \; {\cal I}'_l (\kappa R) -   {\cal I}'_l (\sqrt{\epsilon \mu}\kappa R)  \;{\cal I}_l (\kappa R)  }
{{\cal K}_l (\kappa R) \; {\cal I}'_l ( \sqrt{ \mu/\epsilon}\kappa R) - \sqrt{\epsilon/\mu} \; {\cal I}_l (\sqrt{\epsilon \mu}\kappa R) \; {\cal K}'_l (\kappa R) } \;,\\
T^{\rm EH}_{l} ({\rm i} \xi)&=& T^{\rm HE}_{l} ({\rm i} \xi) = 0 \; ,
\end{eqnarray} 
where $\xi=\kappa c$, ${\cal I}_l (x) = x i_l(x)$ and ${\cal I}'_l(x)= d {\cal I}_l /dx$  with  $i_l(x) = \sqrt{\frac{\pi}{2 x}} I_{l+1/2}(x)$   the modified spherical Bessel function of the first kind.

 The matrix elements of the SSO $\mathbb{K}$ and the operator $\mathbb{M}$ can be easily computed in the  basis of vector spherical harmonics. The corresponding matrices are both diagonal  with respect to multipole indices $(l,m)$ ($-l \le m \le l$), and in addition they are independent of $m$. Therefore, the matrix for  $\mathbb{K}$  has  the structure of $l$-dependent $4 \times 4$  blocks 
$K_{p,r,l,m;q,s,l',m'}= \delta_{ll'}\delta_{mm'} K^{(l)}_{p,r;q,s}$,
where $p,q=E,H$ and $r,s=1,2$ label the  tangential fields $Y_{1,lm}(\hat{r})$ and
$Y_{2,lm}(\hat{r})$ introduced in Eq.~(8.1) of \cite{balian1978}.

 In Figs.~\ref{FigsphereAu}-\ref{Figspherepoly} we show plots of the ratios of the MSE for the CP energy $E^\text{(MSE$_k$)}_\text{CP}$ and the exact result for the CP energy $E^\text{(exact)}_\text{CP}$ obtained from Eq.~(\ref{eq:E_sphere}) versus $d/R$ for Au, Si and polystyrene. In the case of Au, a comparison of  Fig.~\ref{FigsphereAu}(a) with Fig.~\ref{FigsphereAu}(b) shows that the rate of convergence is much faster with the asymmetric choice (C2) of the coefficients.  In fact, with the (C2) choice already ${\rm MSE}_3$ differs from the exact energy by less than one percent  for all displayed separations: specifically, the  maximum error is of $0.6 \%$ for $d/R=1$, while for $d/R=0.03$ the error is as small as $0.1$ \%.  In the case of Si, the performance of the choice (C1) is better than (C2). Indeed, with the (C1) choice the maximum error of  ${\rm MSE}_4$ is of 0.8\%  for $d/R=0.03$ while the minimum error is of 0.2 \% for $d/R=1$, while for  the choice (C2) the maximum error is of 3.4\% for $d/R=1$. In the case of polystyrene, the performance of the (C1) is excellent, since with ${\rm MSE}_3$ the maximum error is of 0.6 \% for $d/R=1$, while for $d/R=0.04$ the error is as low as 0.003 \%.  For polystyrene, the rate of convergence of (C2) is instead very poor.
\begin{table}[H] 
\caption{Oscillator parameteres for Au  \cite{decca2007}\label{tabAu}}
\newcolumntype{C}{>{\centering\arraybackslash}X}
\begin{tabularx}{\textwidth}{CCC}
\toprule
\textbf{ $\omega_j$ (eV$/\hbar$) }	& \textbf{$f_j$ (${\rm eV}^2$$/\hbar^2$)}	& \textbf{$g_j$ (eV$/\hbar$)}\\
\midrule
3.05		& 7.091			& 0.75\\
4.15		& 41.46			& 1.85 \\
5.4		& 2.7			& 1.0\\
8.5		& 154.7			& 7.0 \\
13.5		& 44.55			& 6.0 \\
21.5		& 309.6			& 9.0 \\
\bottomrule
\end{tabularx}
\end{table}

\begin{table}[H] 
\caption{Oscillator parameteres for polystyrene \cite{parsegian} \label{tabpoly}}
\newcolumntype{C}{>{\centering\arraybackslash}X}
\begin{tabularx}{\textwidth}{CCC}
\toprule
\textbf{ $\omega_j$ (eV$/\hbar$) }	& \textbf{$f_j$ (${\rm eV}^2$$/\hbar^2$)}	& \textbf{$g_j$ (eV$/\hbar$)}\\
\midrule
6.35		& 14.6			& 0.65\\
14.0		& 96.9			& 5.0 \\
11.0		& 44.4	         	& 3.5\\
20.1  	& 136.9			& 11.5 \\
\bottomrule
\end{tabularx}
\end{table}

\begin{figure}[H]
\includegraphics[width=1.0\textwidth]{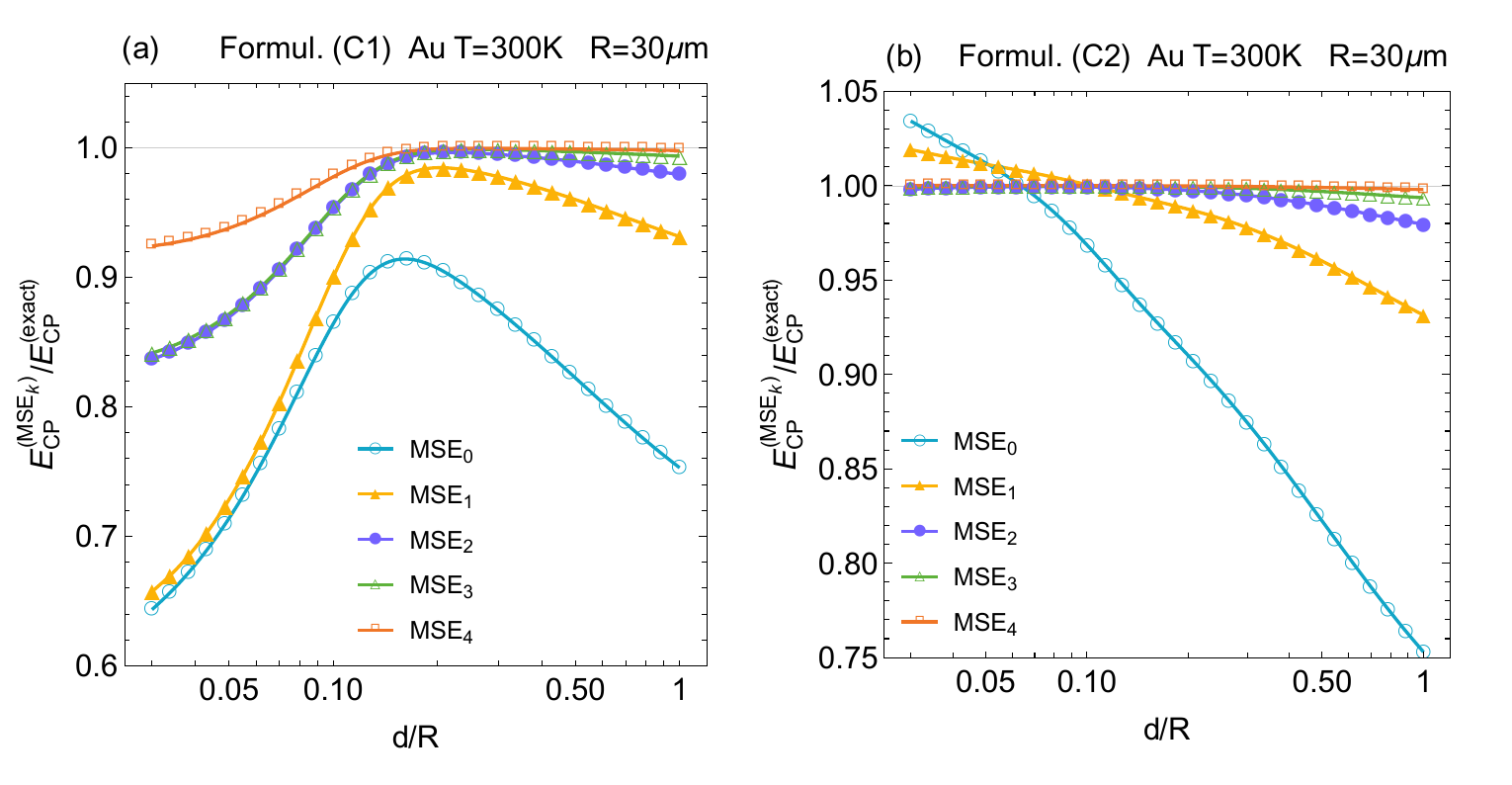}
\caption{MSE of the CP energy for a Au sphere of radius $R=30\; \mu$m at room temperature: (a) for the formulation (C1), (b) for the formulation (C2).  \label{FigsphereAu}}
\end{figure}   
\unskip

\begin{figure}[H]
\includegraphics[width=1.0\textwidth]{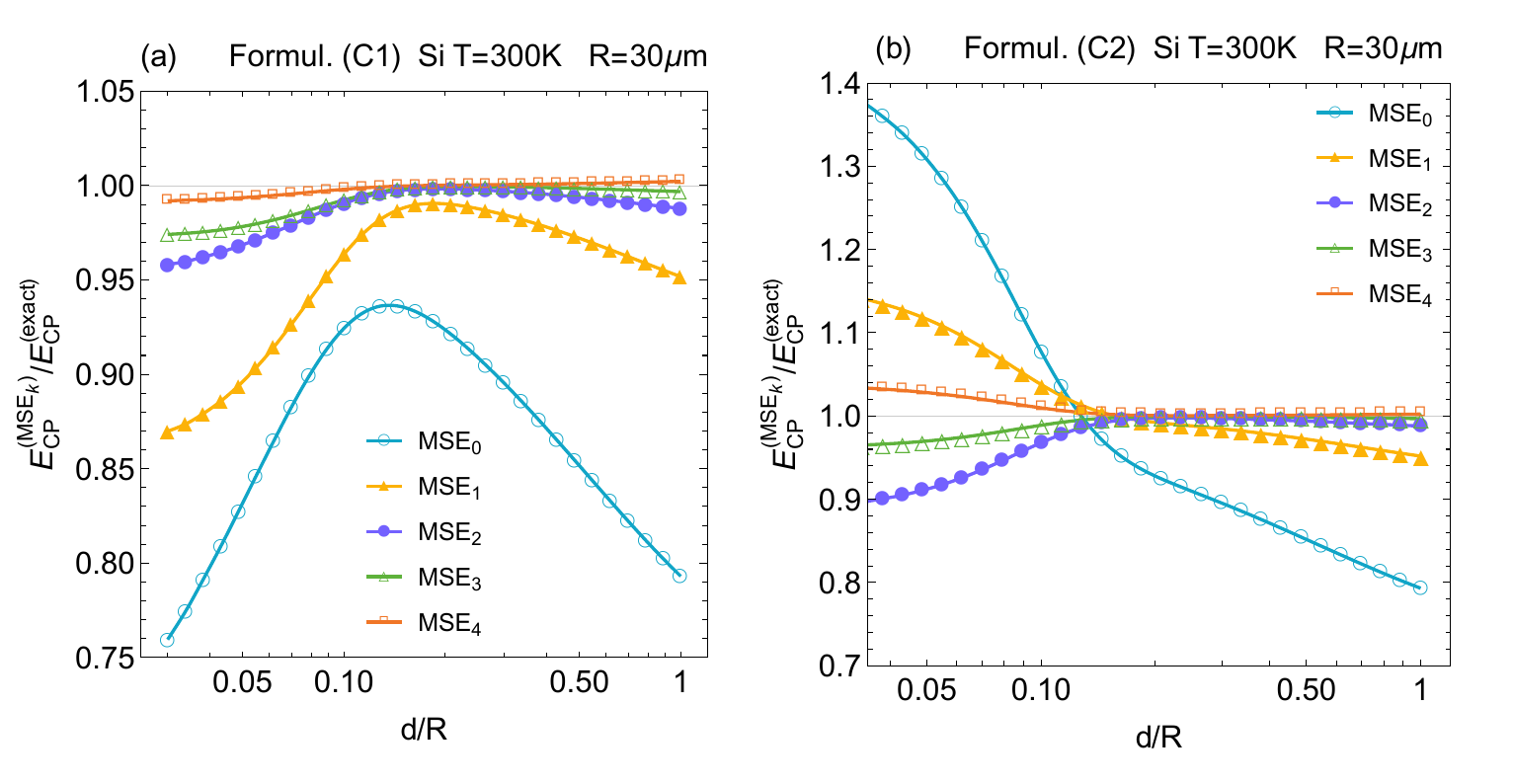}
\caption{MSE of the CP energy for a Si sphere of radius $R=30\; \mu$m at room temperature: (a) for the formulation (C1), (b) for the formulation (C2).\label{FigsphereSi}}
\end{figure}   
\unskip

\begin{figure}[H]
\includegraphics[width=0.5\textwidth]{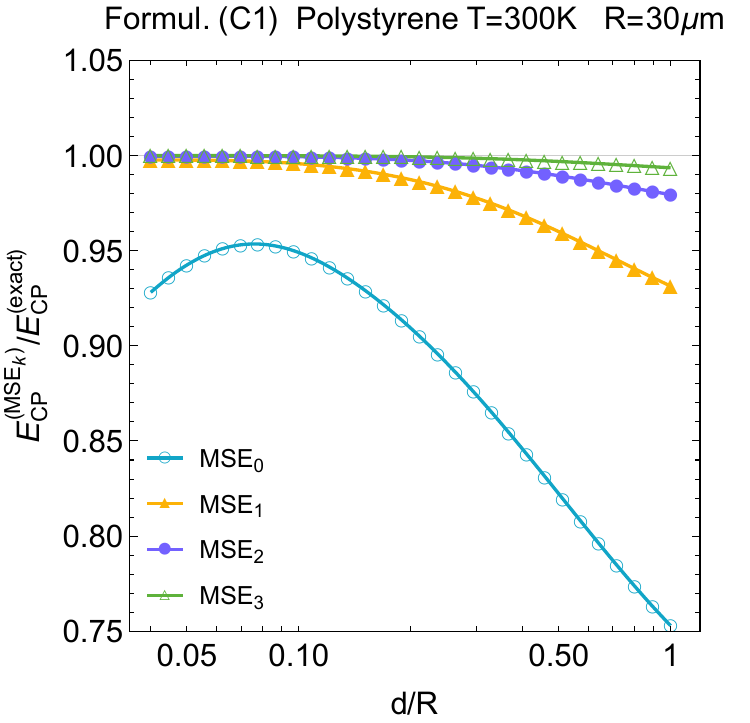}
\caption{MSE of the CP energy for a polystyrene sphere of radius $R=30\; \mu$m at room temperature, for the formulation (C1). \label{Figspherepoly}}
\end{figure}   
\unskip

\subsection{CP energy for a cylinder}

Within the scattering T-matrix approach, the CP interaction energy of a polarizable particle at distance $d$ from the surface of an infinitely long cylinder of radius $R$ with electric and magnetic permeabilities $\epsilon_1=\epsilon$, $\mu_1=\mu$ in vacuum ($\epsilon_0=\mu_0=1$)  is
\begin{eqnarray}
{E}^\text{(exact)}_\text{CP} &=& \frac{k_B T}{\pi} \sideset{}{'}\sum_{n=0}^\infty  \kappa_n^2 \,  \alpha({\rm i} \, \xi_n) 
\int_{-\infty}^\infty dk_z \sum_{m=-\infty}^\infty \\
&\times& \left\{ T^\text{EE}_{k_z m}(i\xi_n) \frac{1}{\kappa_n^2}\left[ k_z^2 {K'_m}^{\!\!2}(p_0 a) + \left( \frac{m^2 k_z^2}{p_0^2 a^2}+p_0^2\right)K_m^2(p_0 a)\right]\right. \nonumber \\
&-& \left. T^\text{HH}_{k_z m}(i\xi_n) \left[ {K'_m}^{\!\!2}(p_0 a) + \frac{m^2}{p_0^2 a^2} K_m^2(p_0 a)\right]\right. \nonumber \\
&+& \left. T^\text{EH}_{k_z m}(i\xi_n) \frac{4 m k_z}{\kappa_n p_0 a} K_m(p_0 a) {K'_m}^{\!\!2}(p_0 a) \right\}
\;, \nonumber
\label{eq:E_cyl}
\end{eqnarray}
where $a=R+d$, $p_0=\sqrt{\kappa^2+k_z^2}$, $m$ is the multipole index, $K_m$ is the modified Bessel function of second kind and $K'_m$ its derivative, and $T^{NM}_{k_z m}$, ($N, M \in \{E,H\}$) are the T-matrix elements of a dielectric cylinder \cite{Noruzifar:2012wk},
\begin{gather}
  \label{eq:T-matrix-elements-mm}
  T^{HH}_{k_z m}(i\xi) = -\frac{I_m(p_0R)}{K_m(p_0 R)} \frac{\Delta_1\Delta_4 +\Upsilon^2}{\Delta_1\Delta_2+\Upsilon^2}\,,\\
  \label{eq:T-matrix-elements-ee}
  T^{EE}_{k_z m}(i\xi)= -\frac{I_m(p_0 R)}{K_m(p_0 R)} \frac{\Delta_2\Delta_3 +\Upsilon^2}{\Delta_1\Delta_2+\Upsilon^2}\,,\\
  \label{eq:T-matrix-elements-me}
  T^{HE}_{k_z m}(i\xi) = -T^{EH}_{k_z m}(i\xi) =  \frac{\Upsilon}{ \sqrt{\epsilon\mu} (p_0 R)^2 K_m(p_0 R)^2}
\frac{1}{\Delta_1\Delta_2 +\Upsilon^2} \,,
\end{gather}
with $I_m$ the modified Bessel function of first kind, and 
\begin{equation}
  \label{eq:def_K}
  \Upsilon = \frac{m k_z}{\sqrt{\epsilon\mu} R^2 \kappa} \left( \frac{1}{p^2} - \frac{1}{p_0^2}\right)\,,
\end{equation}
with $p=\sqrt{\epsilon\mu \kappa^2+k_z^2}$ and
\begin{eqnarray}
  \label{eq:def_Delats}
  \Delta_1 &= & \frac{I'_m(pR)}{p R I_m(pR)} -\frac{1}{\epsilon} \frac{K'_m(p_0R)}{p_0R K_m(p_0R)}\,,\\
 \Delta_2 &= & \frac{I'_m(pR)}{p R I_m(pR)} -\frac{1}{\mu} \frac{K'_m(p_0R)}{p_0R K_m(p_0R)}\,,\\
 \Delta_3 &= & \frac{I'_m(pR)}{p R I_m(pR)} -\frac{1}{\epsilon} \frac{I'_m(p_0R)}{p_0R I_m(p_0R)}\,,\\
 \Delta_4 &= & \frac{I'_m(pR)}{p R I_m(pR)} -\frac{1}{\mu} \frac{I'_m(p_0R)}{p_0R I_m(p_0R)} \, .
\end{eqnarray}
Notice that in general the polarization is {\it not conserved} under scattering, i.e.,
$T^{EH}_{k_z m} \neq 0 \neq T^{HE}_{k_z m}$. This property, together with its quasi-2D shape, makes the cylinder an important benchmark test for the convergence of the MSE.

\begin{figure}[H]
\includegraphics[width=1.0\textwidth]{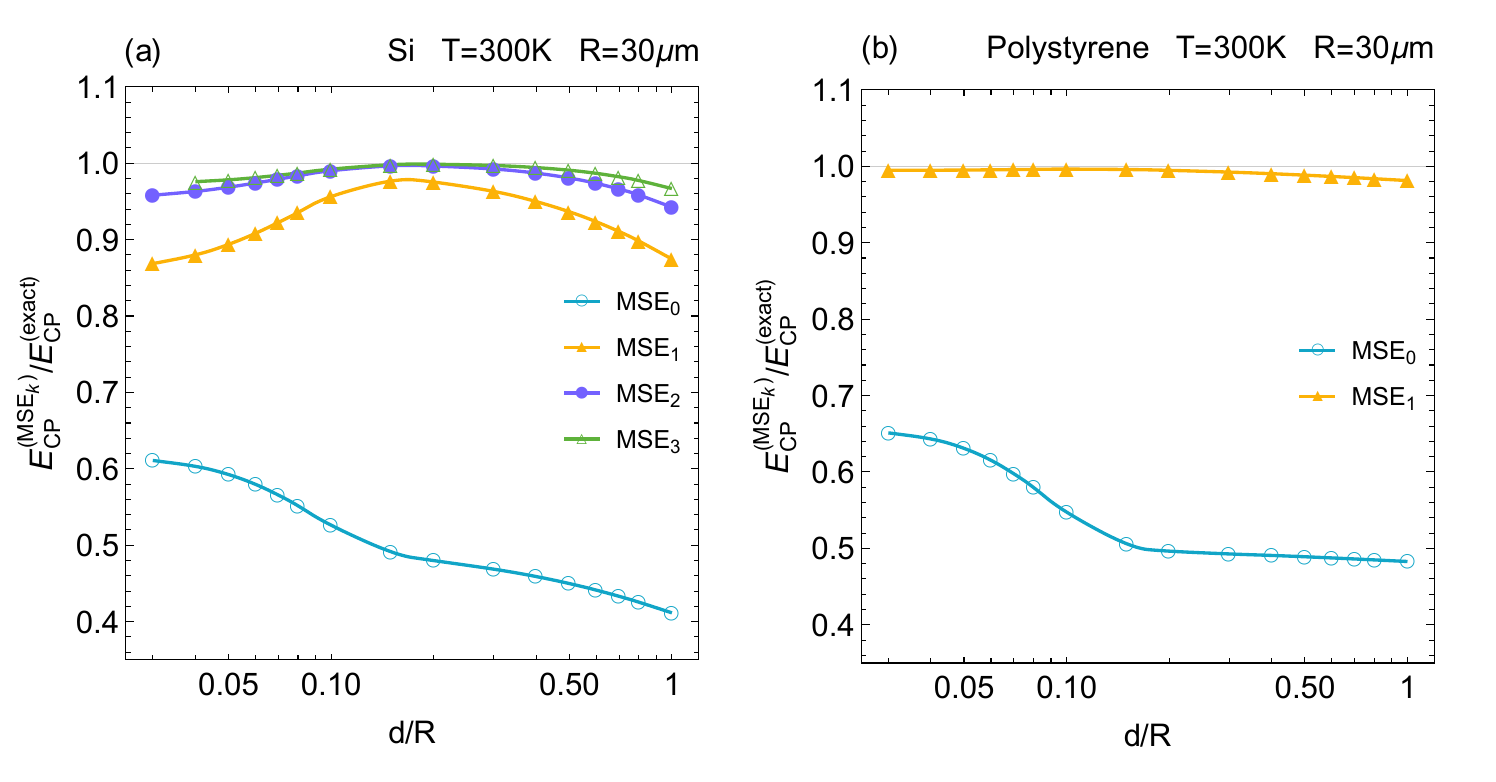}
\caption{MSE of the CP energy for (a) a Silicon cylinder and (b) a polystyrene cylinder of radius $R=30\; \mu$m at room temperature $T=300K$, for the formulation (C1). \label{fig:5}}
\end{figure}

The CP energy can be easily obtained as a MSE since the SSO $\mathbb{K}$ and the operator $\mathbb{M}$ can be computed by substituting for the free Green functions in Eqs.~(\ref{eq:1}), (\ref{eq:3}) an expansion in vector cylindrical waves.
In Fig.~\ref{fig:5} we show again numerical results for the ratio of the MSE for the CP energy $E^\text{(MSE$_k$)}_\text{CP}$ at MSE order $k$ and the exact result for the CP energy $E^\text{(exact)}_\text{CP}$ obtained from Eq.~(\ref{eq:E_cyl}). 
The materials, temperature and geometric lengths are the same as in the case of a sphere. For Si we observe that the MSE with choice (C1) has converged at order $\text{MSE}_{3}$ to the exact energy within about $3 \%$, with the largest deviations at the shortest ($2.4 \%$) and longest considered ($3.3 \%$) separation. The deviation is minimal at intermediate distances around $d/R=0.2$ with an error of only $0.1 \%$. Hence the performance of the MSE for an infinite cylinder is very similar to a compact sphere. We did not consider the coefficients (C2) as they performed worse than choice (C1) for a sphere. For polystyrene we consider again only the choice (C1), for the same reason. Due to its low dielectric contrast, we expect the 
choice (C1) to give excellent convergence of the MSE at low order. Indeed, the rate of convergence is so fast that the MSE can be terminated at order $\text{MSE}_{1}$ already, with a maximum deviation from the exact energy of only $1.9\%$ at the separation $d=R$. In general, we note that with the choice (C1) the lowest order $\text{MSE}_{0}$ the estimate of the energy for the cylinder is less good than for the sphere. This is presumably due to arbitrarily long range charge and current fluctuations along the cylinder  which require at least one power the operator $\mathbb{K}$ to be described properly.

Finally, it is important to discuss the case of a metal, like Au. As the dielectric function diverges in the limit $\kappa\to 0$, the classical term $n=0$ of the Matsubara sum resembles that of a perfect conductor. We had shown that for a cylinder the SSO $\mathbb{K}$, for the choice (C1), in the partial wave channel $m=0$ has an eigenvalue that approaches unity when $\kappa\to 0$ and $\epsilon\to\infty$ \cite{bimonte2023}. For the choice (C2) the situation is even worse as there is an eigenvalue approaching unity in all partial wave channels. We expect this property to persist for all quasi-2D shapes with a compact cross section. Hence, for such metallic shapes the classical term $n=0$ cannot be obtained from a MSE. 
However, our 
surface scattering approach is also useful for zero frequency $\kappa=0$  as the inverse of $\mathbb{M}= \mathbb{I}-\mathbb{K}$ can be computed directly, without resorting to a MSE. We note that for $\kappa=0$ the expression for $\mathbb{K}$ simplifies considerably, in particular in the perfect conductor limit \cite{bimonte2023}. 

To conclude, we have demonstrated that the MSE provides an excellent device to compute Casimir-Polder interactions with high precision for a wide range of materials. We stress that this conclusion is not specific to the shapes considered here but is expected to hold generically for any compact 3D shape or quasi-2D shape. Here we considered a sphere and a cylinder only for the reason that for those shapes exact results are known and hence the convergence of our MSE can be tested. Most importantly, for general shapes where the T-matrix is not known, the SSO ${\mathbb K}$ can be computed and the MSE implemented to obtain high precision results for the interaction.

\unskip

\end{document}